\documentclass[twocolumn,times]{aastex631}

\usepackage{multirow}
\usepackage{amsmath}

\begin{document}

\title{Optical and Near-infrared Observations of the Distant but Bright `New Year's Burst' GRB\,220101A} 

\correspondingauthor{Wei-Hua Lei, Dong Xu}
\email{leiwh@hust.edu.cn, dxu@nao.cas.cn}

\author[0000-0002-9022-1928]{Zi-Pei Zhu}
\affiliation{Department of Astronomy, School of Physics, Huazhong University of Science and Technology, Wuhan, 430074, China}
\affiliation{Key Laboratory of Space Astronomy and Technology, National Astronomical Observatories, Chinese Academy of Sciences, Beijing, 100101, China}

\author[0000-0003-3440-1526]{Wei-Hua Lei}
\affiliation{Department of Astronomy, School of Physics, Huazhong University of Science and Technology, Wuhan, 430074, China}

\author[0000-0002-7517-326X]{Daniele B. Malesani}
\affiliation{Niels Bohr Institute, University of Copenhagen, Jagtvej 155, 2200, Copenhagen N, Denmark}
\affiliation{Department of Astrophysics/IMAPP, Radboud University, PO Box 9010, 6500 GL, The Netherlands}

\author{Shao-Yu Fu}
\affiliation{Key Laboratory of Space Astronomy and Technology, National Astronomical Observatories, Chinese Academy of Sciences, Beijing, 100101, China}
\affiliation{School of Astronomy and Space Science, University of Chinese Academy of Sciences, Chinese Academy of Sciences, Beijing 100049, China}

\author{Dong-Jie Liu}
\affiliation{Department of Astronomy, School of Physics, Huazhong University of Science and Technology, Wuhan, 430074, China}

\author[0000-0003-3257-9435]{Dong Xu}
\affiliation{Key Laboratory of Space Astronomy and Technology, National Astronomical Observatories, Chinese Academy of Sciences, Beijing, 100101, China}

\author{Paolo D'Avanzo}
\affiliation{Brera Astronomical Observatory, via Bianchi 46, I-23807, Merate (LC), Italy}

\author[0000-0001-6991-7616]{José Feliciano Agüí Fernández}
\affiliation{Instituto de Astrof\'isica de Andaluc\'ia, Glorieta de la Astronom\'ia s/n, 18008 Granada, Spain.}

\author[0000-0002-8149-8298]{Johan P. U. Fynbo}
\affiliation{The Cosmic Dawn Centre (DAWN)}
\affiliation{Niels Bohr Institute, University of Copenhagen, Jagtvej 155, 2200, Copenhagen N, Denmark}

\author{Xing Gao}
\affiliation{Xinjiang Astronomical Observatory, Chinese Academy of Sciences, Urumqi,  Xinjiang 830011, China}

\author{Ana Nicuesa Guelbenzu}
\affiliation{Th\"uringer Landessternwarte Tautenburg, 07778 Tautenburg, Germany}

\author{Shuai-Qing Jiang}
\affiliation{Key Laboratory of Space Astronomy and Technology, National Astronomical Observatories, Chinese Academy of Sciences, Beijing, 100101, China}
\affiliation{School of Astronomy and Space Science, University of Chinese Academy of Sciences, Chinese Academy of Sciences, Beijing 100049, China}

\author[0000-0003-2902-3583]{David Alexander Kann}
\affiliation{Instituto de Astrof\'isica de Andaluc\'ia, Glorieta de la Astronom\'ia s/n, 18008 Granada, Spain.}
\affiliation{Hessian Research Cluster ELEMENTS, Giersch Science Center, Max-von-Laue-Strasse 12, Goethe University Frankfurt, Campus Riedberg, 60438 Frankfurt am Main, Germany}

\author{Sylvio Klose}
\affiliation{Th\"uringer Landessternwarte Tautenburg, 07778 Tautenburg, Germany}

\author[0000-0002-7420-6744]{Jin-Zhong Liu}
\affiliation{Xinjiang Astronomical Observatory, Chinese Academy of Sciences, Urumqi,  Xinjiang 830011, China}
\affiliation{Key Laboratory of Space Astronomy and Technology, National Astronomical Observatories, Chinese Academy of Sciences, Beijing, 100101, China}

\author{Xing Liu}
\affiliation{Key Laboratory of Space Astronomy and Technology, National Astronomical Observatories, Chinese Academy of Sciences, Beijing, 100101, China}
\affiliation{Key Laboratory of Cosmic Rays, Ministry of Education, Tibet University, Lhasa, Tibet 850000, China}

\author[0000-0002-4036-7419]{Massimiliano De Pasquale}
\affiliation{University of Messina, Mathematics, Informatics, Physics and Earth Science Department, Via F.D. D’Alcontres 31, Polo Papardo,
98166, Messina, Italy}

\author[0000-0001-7717-5085]{Antonio de Ugarte Postigo}
\affiliation{Université Côte d'Azur, Observatoire de la Côte d'Azur, Artemis, CNRS, 06304 Nice, France}

\author{Bringfried Stecklum}
\affiliation{Th\"uringer Landessternwarte Tautenburg, 07778 Tautenburg, Germany}

\author[0000-0002-7978-7648]{Christina Th\"{o}ne}
\affiliation{Astronomical Institute of the Czech Academy of Sciences (ASU-CAS), Fri\v
cova 298, 251 65 Ond\v rejov, CZ}

\author{Joonas Kari Markku Viuho}
\affiliation{The Cosmic Dawn Centre (DAWN)}
\affiliation{Niels Bohr Institute, University of Copenhagen, Jagtvej 155, 2200, Copenhagen N, Denmark}

\author{Yi-Nan Zhu}
\affiliation{CAS Key Laboratory of Optical Astronomy, National Astronomical Observatories, Chinese Academy of Sciences,Beijing 100101, China}

\author{Jing-Da Li}
\affiliation{Department of Astronomy, Beijing Normal University, Beijing 100875, China}

\author[0000-0002-3100-6558]{He Gao}
\affiliation{Department of Astronomy, Beijing Normal University, Beijing 100875, China}

\author{Tian-Hua Lu}
\affiliation{Key Laboratory of Space Astronomy and Technology, National Astronomical Observatories, Chinese Academy of Sciences, Beijing, 100101, China}
\affiliation{School of Astronomy and Space Science, University of Chinese Academy of Sciences, Chinese Academy of Sciences, Beijing 100049, China}

\author{Shuo Xiao}
\affiliation{Key Laboratory of Particle Astrophysics, Institute of High Energy Physics, Chinese Academy of Sciences, 19B Yuquan Road, Beijing 100049, China}

\author[0000-0002-5400-3261]{Yuan-Chuan Zou}
\affiliation{Department of Astronomy, School of Physics, Huazhong University of Science and Technology, Wuhan, 430074, China}

\author[0000-0002-9422-3437]{Li-Ping Xin}
\affiliation{Key Laboratory of Space Astronomy and Technology, National Astronomical Observatories, Chinese Academy of Sciences, Beijing, 100101, China}

\author{Jian-Yan Wei}
\affiliation{Key Laboratory of Space Astronomy and Technology, National Astronomical Observatories, Chinese Academy of Sciences, Beijing, 100101, China}
\affiliation{School of Astronomy and Space Science, University of Chinese Academy of Sciences, Chinese Academy of Sciences, Beijing 100049, China}

\begin{abstract}

High-redshift gamma-ray bursts (GRBs) provide a powerful tool to probe the early universe, but still for relatively few do we have good observations of the afterglow. We here report the optical and near-infrared observations of the afterglow of a relatively high-redshift event, GRB\,220101A, triggered on New Year's Day of 2022. With the optical spectra obtained at XL2.16/BFOSC and NOT/ALFOSC, we determine the redshift of the burst at $z= 4.615$. Based on our optical and near-infrared data, combined with the X-ray data, we perform multiband fit with the python package \emph{afterglowpy}. A jet-break at $\sim$ 0.7 day post-burst is found to constrain the opening angle of the jet as $\sim$ 3.4 degree. We also determine circumburst density of $n_0 = 0.15\ {\rm  cm}^{-3}$ as well as kinetic energy $E_{\rm K, iso} = 3.52\times 10^{54}$ erg. The optical afterglow is among the most luminous ever detected. We also find a ``mirror'' feature in the lightcurve during the prompt phase of the burst from 80 s to 120 s. The physical origin of such mirror feature is unclear.

\end{abstract}

\keywords{\href{http://astrothesaurus.org/uat/629}{Gamma-ray bursts}}

\section{Introduction} \label{sec:intro}

Gamma-ray bursts (GRBs) are the most energetic and luminous transient events in the universe. The duration of GRBs usually ranges from sub-seconds to several hundred seconds. Based on the statistic of prompt emission $T_{90}$ duration timescale and the spectral hardness of the bursts, GRBs can generally be divided into two categories, i.e., short bursts with $T_{90} < 2$ s and hard spectral, which is confirmed by the origin of neutron star mergers and associated with a kilonovae \citep{1993ApJ...413L.101K, 2009ApJ...703.1696Z, 2017ApJ...848L..12A}. Long bursts with $T_{90} > 2$ s and soft spectral, which is confirmed the origin of the collapse of massive stars and associated with broad-lined Type Ic supernovae \citep{Galama+1999,2006ARA&A..44..507W}. There are some confusing exceptions \citep[e.g.,][]{2006Natur.444.1050D, 2006Natur.444.1047F,2006Natur.444.1044G,2022Natur.612..223R}

Since the first afterglow counterpart discovered by \textit{BeppoSAX} in the 1990s, the redshift was measured for the first time and the cosmological origin of GRBs was determined \citep[see][for a review]{2018pgrb.book.....Z}. With the Neil Gehrels \textit{Swift} Observatory (\emph{Swift} hereafter) launched in 2004, more and more bursts are detected with accurate position \citep{2004ApJ...611.1005G}. Nearly 600 bursts redshifts\footnote{\url{https://www.mpe.mpg.de/~jcg/grbgen.html}} have been measured so far, and they range from 0 to $\sim$ 9.4 \citep{2011ApJ...736....7C}. The typical isotropic equivalent energy of GRBs ranges between $10^{50}-10^{54}$ ergs within a short duration \citep{2017ApJ...837..119A}. Most bursts with measured redshift are in the redshift range from 0 to 3, and only relatively few have measured redshift larger than 4 \citep{2022JApA...43...82G}. High-redshift GRBs, however, are an important complementary probe of the early universe, and they could potentially point us to the mysterious Population III stars \citep{2000ApJ...536....1L,2015ApJ...808..139S,2015JHEAp...7...35S,2022ApJ...929..111F}. In the past 20 years, about one high-redshift (i.e., $z\gtrsim4$) GRB per year was detected on average. Therefore, the detection of high-\textit{z} GRB is still of significant interest. For a single event, multiband observations are important to investigate the intrinsic properties of these high-\textit{z} bursts. This, on the other hand, is great help to study whether such burst is markedly different from other GRBs at lower redshifts.

Recently, a relatively high-redshift and long duration gamma-ray burst GRB\,220101A.
The redshift of the burst is first measured by Xinglong 2.16m Telescope \citep{2022GCN.31353....1F}, and then confirmed by NOT \citep{2022GCN.31359....1F} and Copernico Telescope \citep{2022GCN.31363....1T}. AGILE observations of GRB\,220101A took place at the beginning of 2022 January 1. It is the second-farthest event detected by AGILE and was recognized as one of the most energetic gamma-ray bursts (GRBs) ever detected since their discovery \citep{Ursi2022}. \citet{Mei2022} gave joint multiband analysis from soft X-rays to high energies (up to $\sim 1$ GeV) of GRB\,220101A. \citet{Jin2023} carried out time-resolved analysis of the Swit/UVOT 150s exposure of GRB\,220101A, and found a rapidly-evolving optical/ultraviolet flare with an unprecedented-high absolute AB magnitude $\sim -39.4$. 

Here we report our optical and near-infrared observations of GRB\,220101A. For our modelling, we make use of the \emph{afterglowpy} python package \citep{Ryan2020}. \emph{Afterglowpy} is a publicly available open-source Python package for the numerical computation of structured jet afterglows. This paper is organized as follows: we describe our multiband observations and redshift measurements in Section \ref{sec:obs}. The combined analysis of multiband data with X-ray light curve are presented in Section \ref{sec:model}. We also discuss results and summary the conclusions in Section \ref{sec:discussion} and Section \ref{sec:summary}, respectively. A standard cosmology model is adopted with $H_{0}=67.3\ \rm{km s^{-1} Mpc^{-1}}$, $\Omega_{M}$=0.315, $\Omega_{\Lambda}$=0.685 \citep{Planck+2014}.

\section{Observations and data reduction} \label{sec:obs}

GRB\,220101A first triggered Neil Gehrels \emph{Swift} Observatory (short as \emph{Swift}) Burst Alert Telescope (BAT, \citealt{2005SSRv..120..143B}) at 05:09:55 UT on Jan 1st 2022 \citep{2022GCN.31347....1T}, also triggered other high energy satellites like Fermi/GBM \citep{2022GCN.31360....1L}, Fermi/LAT \citep{2022GCN.31350....1A}, AGILE \citep{2022GCN.31354....1U} and Konus-Wind \citep{2022GCN.31433....1T}. The $\rm T_{90}$ ($15-350$ keV) of the burst is $173.36 \pm 12.76$ s \citep{2022GCN.31369....1M}. X-ray Telescope (XRT, \citealt{2005SSRv..120..143B}) started observations  80.8 s  after the BAT trigger and found a bright, uncatalogued X-ray source within the BAT error circle \citep{2022GCN.31349....1O}. About 9 s later, Ultra-Violet/Optical Telescope (UVOT, \citealt{2005SSRv..120...95R}) found a source with a \emph{white} band brightness of 14.7 at coordinate: 
R.A., decl. (J2000) = $00^h05^m24.80^s,+31^{\circ}46^{'}08.4^{''}$, 
which is consistent with the XRT position \citep{2022GCN.31351....1K}. 
We adopted the analysis results of the XRT repository produces \citep{Evans+2007,Evans+2009} and downloaded the $0.3-10$ keV unabsorbed light curve from the UK \emph{Swift} Science Data Centre\footnote{\url{https://www.swift.ac.uk/xrt\_curves/}}.
In our analysis of the afterglow, we excluded the XRT data within 10 ks since the BAT trigger.
\cite{Jin2023} analyzed the UVOT data, found the early light curve showing a rapidly-evolving flare. 
In order to show the full light curve of the burst, we collected the \emph{white} band data from the paper mentioned above. 
We also noticed the mirror feature during the prompt emission, and downloaded the time-tagged event (TTE) data from Fermi Science Support Center's FTP site\footnote{\url{https://heasarc.gsfc.nasa.gov/FTP/fermi/data/gbm/bursts/}}. We selected the NaI detectors n6 and n7 that were triggered by the GRB. The processing of the data is done with the GBM Data Tools \footnote{\url{https://fermi.gsfc.nasa.gov/ssc/data/analysis/gbm/}}. We extracted the 64 ms resolution counts data of both detectors and merged them as the light curve data.

\subsection{HST observations}

Hubble Space Telescope (HST) observed the source on 7th Feb. 2022 with F125W and F775W filter equipped on the Wide-Field Camera 3 (Proposal ID: 16838, PI: Andrew S. Fruchter). We obtained the publicly available data from Mikulski Archive for Space Telescopes (MAST) and measured the photometry by \textit{photutils} \citep{2021zndo...5525286B} with PSF model constructed by \textit{PSFEx} \citep{2011ASPC..442..435B}. Our photometric results are presented in Table~\ref{tab:phot}. The F775W and F125W filters are treated as \emph{i} and \emph{J} filters in the following analysis, respectively, for their similar wavelength.

\subsection{Ground-based optical and near-infrared observations}

\begin{figure*}[htp]
\center
\includegraphics[width=0.70\textwidth]{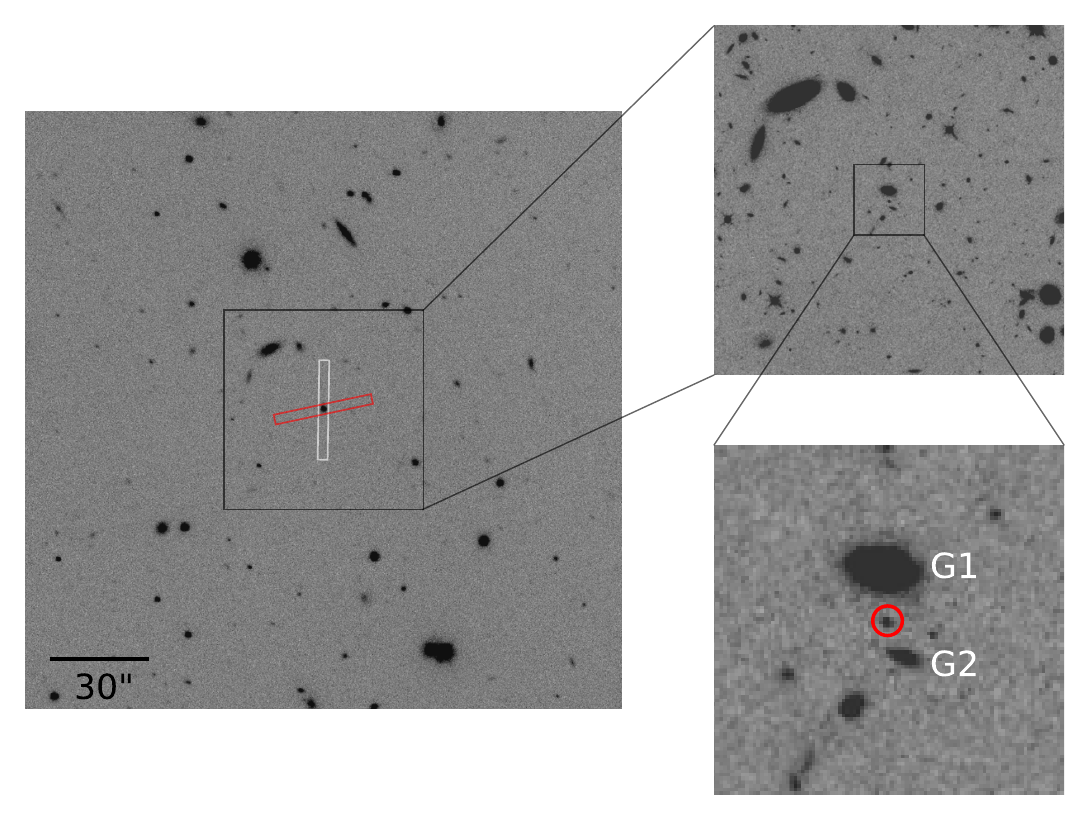}
\caption{The \emph{r} band field view of GRB\,220101A obtained by ALFOSC on the first night (left) and the later time observations with \emph{HST} in F125W filter (top and bottom right). 
The slit direction of ALFOSC and BFOSC is also shown in the left panel of the figure, colored in red and white rectangle, respectively.
In the zoomed view of \emph{HST}, the burst is circled in red and two galaxies nearby, named G1 (R.A., decl. (J2000) = 00:05:24.83, +31:46:09.91) and G2 (R.A., decl. (J2000) = 00:05:24.77, +31:46:06.86). The burst offset to G1 and G2 is $1\farcs829$ and $1\farcs377$, corresponding to 12.2 and 9.2 kpc at a common redshift of 4.615, respectively. Upper north and left east.} 
\label{fig:location}
\end{figure*}

We performed our multiband photometric follow-up with the following factories: the Beijing Faint Object Spectrograph and Camera (BFOSC) on the Xinglong 2.16m Telescope \citep{Fan+2016}, the Ningbo Bureau Of Education And Xinjiang Observatory Telescope (NEXT; 0.6 m located at Xingming Observatory, China), the Alhambra Faint Object Spectrograph and Camera (ALFOSC) on the Nordic Optical Telescope (NOT; 2.56 m at the Roque de los Muchachos observatory, La Palma, Spain), the Calar Alto Faint Object Spectrograph (CAFOS) on the Centro Astron\'{o}mico Hispano en Andaluc\'{i}a (CAHA) 2.2 m telescope located at Calar Alto, Spain, the Near Infrared Camera Spectrometer (NICS) on the Telescopio Nazionale Galileo (TNG; 3.58 m telescope located on the Island of San Miguel del La Palma in the Canary Islands), the TAUtenburg KAMera (TAUKAM) $\rm 6k\times6k$ CCD on the Tautenburg 1.34 m Schmidt Telescope \citep{2016SPIE.9908E..4US}, the Espectr\'{o}grafo Multiobjeto Infra-Rojo(EMIR) on the Gran Telescopio CANARIAS (GTC; 10.4 m telescope). The celestial location of the burst is shown in Fig.~\ref{fig:location}.

\begin{figure*}[htp]
\center
\includegraphics[width=0.95\textwidth]{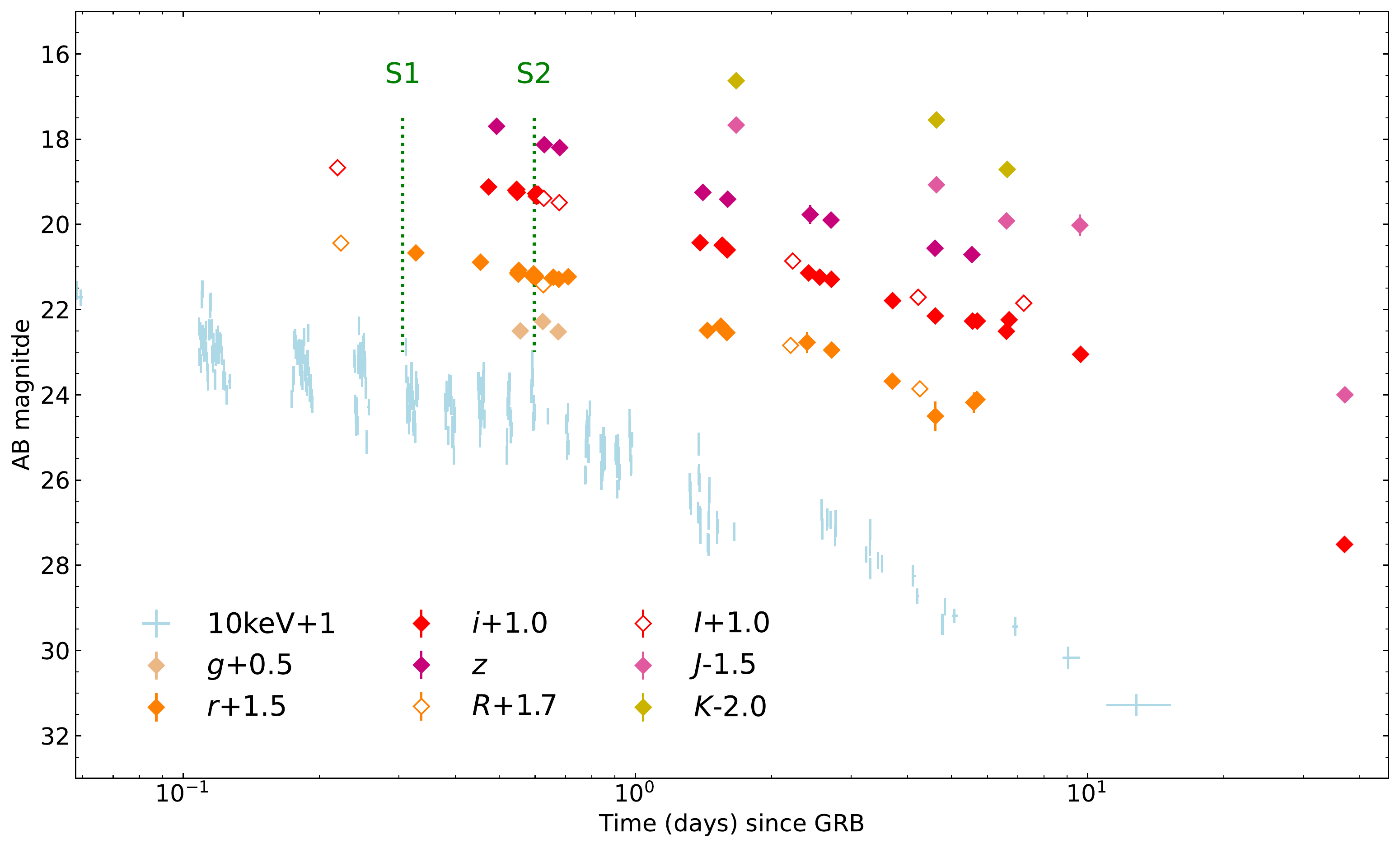}
\caption{The multiband light curve of GRB\,220101A. The points in the figure are in the AB system and have been corrected for Galactic extinction, which is E(B-V) = 0.05 \citep{2011ApJ...737..103S}.
The dotted lines labeled with S1 and S2 represent the beginning time of spectroscopic observation of BFOSC and ALFOSC. The last \textit{i} and \textit{J} band points correspond to the HST F775W and F125W detection, respectively.}
\label{fig:lc}
\end{figure*}

After standard data reduction with {\em IRAF} \citep{Tody+1986} and astrometric calibrated by {\em Astrometry.net} \citep{Lang+2010}, the apparent photometric were calibrated with the {\em Sloan Digital Sky Survey} ({\em SDSS}) 14th data release \citep{Abolfathi+2018} while the near-infrared data were calibrated with the 2MASS\footnote{\url{https://irsa.ipac.caltech.edu/Missions/2mass.html}} catalogue \citep{2006AJ....131.1163S}. The {\em Johnson-Cousin} filters are calibrated with the converted magnitude from the {\em Sloan} system\footnote{\url{https://www.sdss.org/dr12/algorithms/sdssUBVRITransform/\#Lupton}}.
The details of the filters observed by these factories and the photometric results are presented in Table~\ref{tab:phot} and shown in Fig.~\ref{fig:lc}.

Since the discovery of the afterglow was reported, many ground-based observatories contributed to the observation of the burst and published their results on the GRB Coordinates Network (GCN)\footnote{\url{https://gcn.gsfc.nasa.gov/gcn3_archive.html}}. We also include in our analysis the observations reported by the Liverpool Telescope team. These data are calibrated with \textit{SDSS} catalogue and reported in AB magnitudes\citep{2022GCN.31357....1P,2022GCN.31425....1P}.

\begin{figure}[htp]
\center
\includegraphics[width=0.45\textwidth]{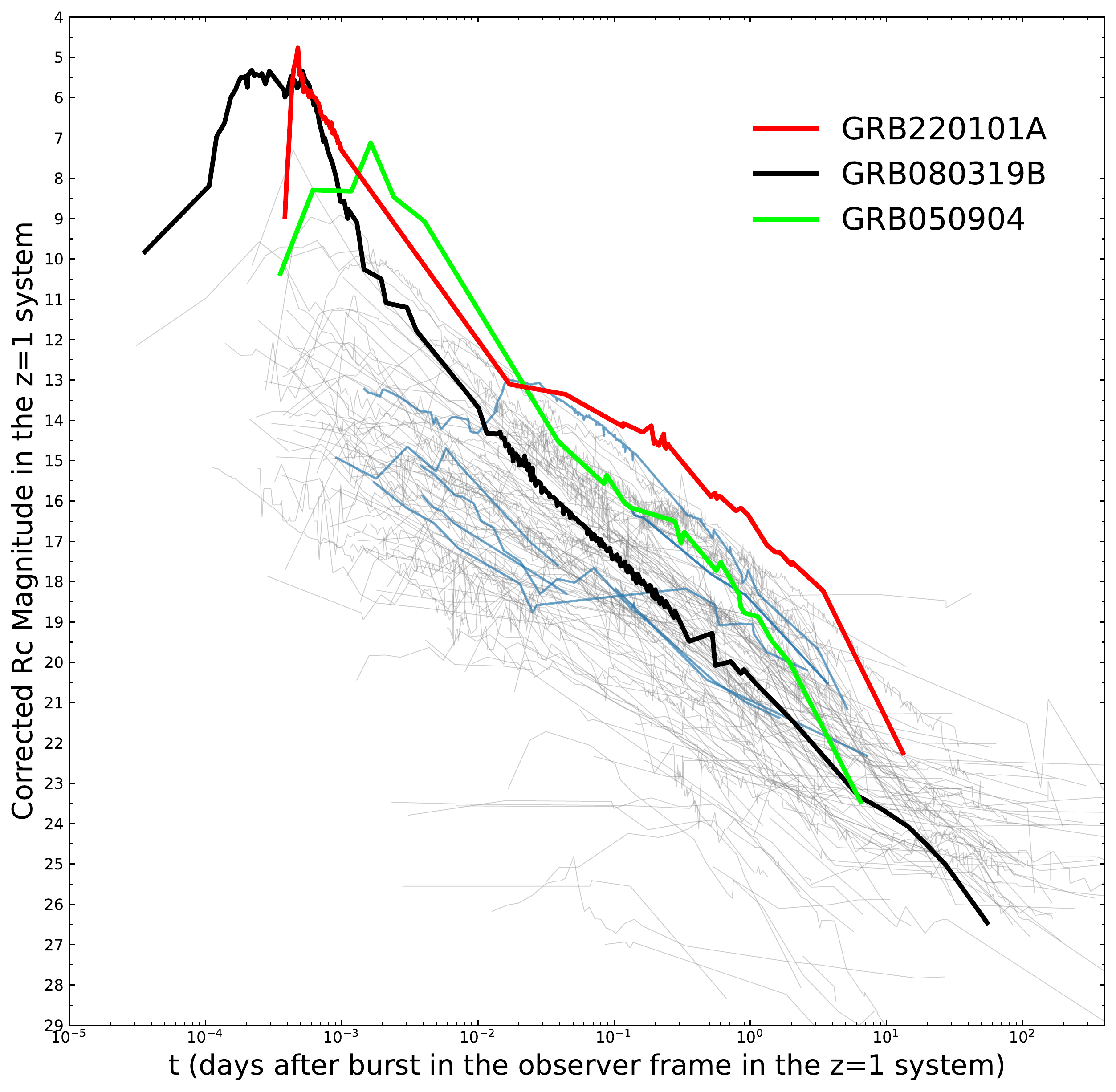}
\caption{`Kann plot' comparing a large sample of GRB optical afterglow light curves shifted in time and flux to a common redshift of $z = 1$ following \citet{Kann2010}. The gray background are historical data of other GRB light curves. GRB\,220101A is shown as a red solid line, which lies at the top of the distribution in terms of luminosity. The bursts of redshift larger than 4 are also colored with light blue.}
\label{fig:lc_comparision}
\end{figure}

In order to place the afterglow of GRB\,220101A in the context of the afterglow sample, we shifted (in time and flux) the afterglow to a redshift $z = 1$. The complete light curve is shown in comparison with other afterglows in Fig. \ref{fig:lc_comparision}. It is immediately apparent that the optical afterglow of GRB\,220101A was intrinsically very bright. It ranks among the brightest afterglows at the epoch of one day.

\subsection{Spectroscopy}

\begin{figure*}
	\centering
	\begin{minipage}{0.95\linewidth}
		\centering
		\includegraphics[width=1\textwidth]{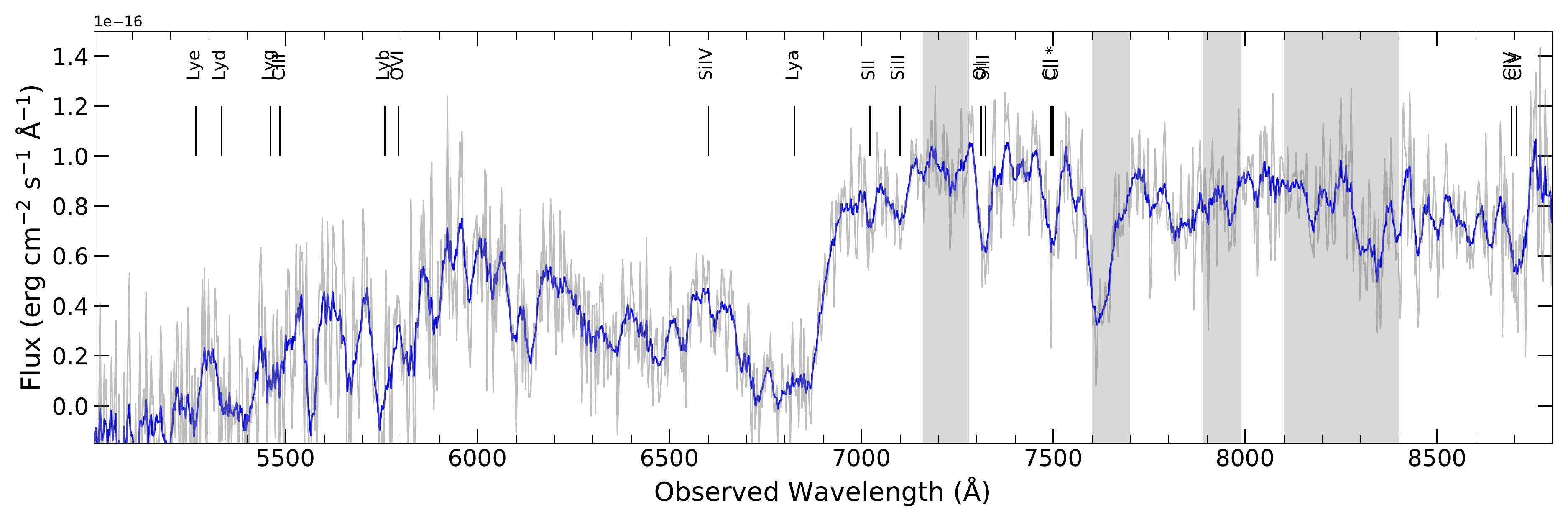}
	\end{minipage}
	\begin{minipage}{0.95\linewidth}
		\centering
		\includegraphics[width=1\textwidth]{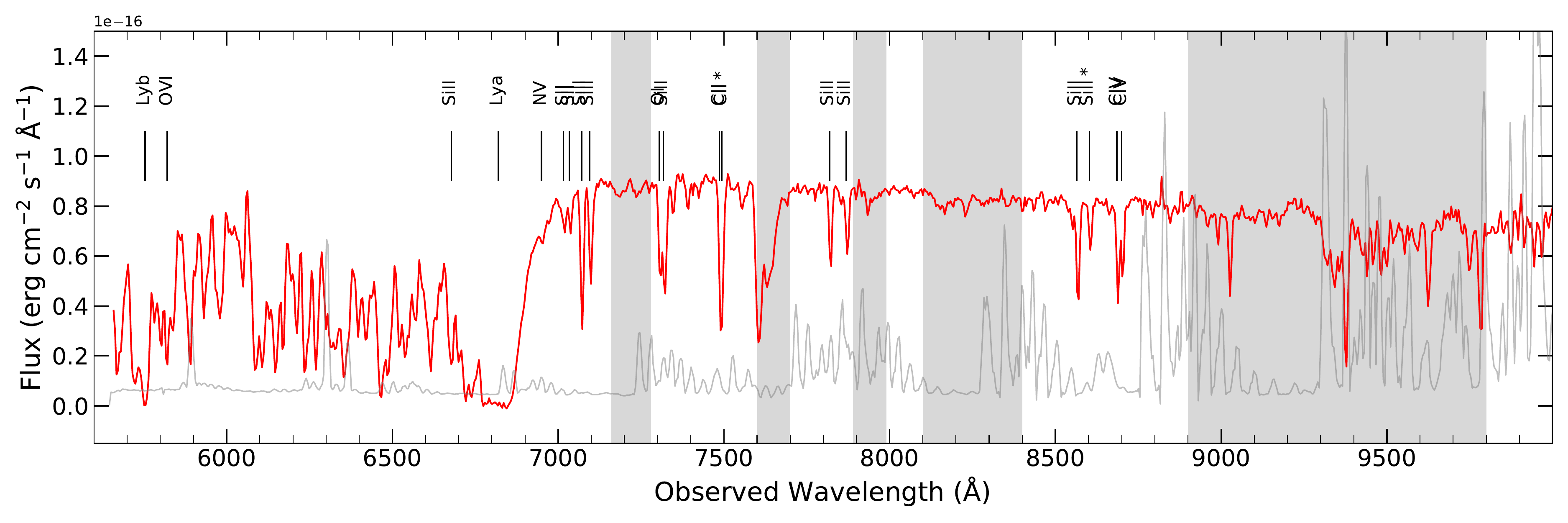}
	\end{minipage}
\caption {
The spectra of afterglow obtained by Xinglong 2.16m/BFOSC and NOT/ALFOSC.  
\textbf{Top panel:} The spectrum  obtained by BFOSC. The gray line is the raw spectrum and the blue is the smoothed for display purpose.
\textbf{Bottom panel:} The spectrum obtained by ALFOSC. The gray line is the background spectrum of sky and the red line is the raw spectrum of the burst.  
In the both panels, the identified metal absorption lines are indicated with vertical lines in the figure, and the possible lines at the left of $\rm Ly\alpha$ are also marked.
The gray vertical lines indicate the telluric features in both panels.
} 
\label{fig:spec}
\end{figure*}

After we obtained the multiband (\textit{BVRI}) photometry results of BFOSC, we noticed that a very significant drop between \textit{B} and \textit{V} band with $ m_B-m_V > 2$, and subsequently realized that this could be a possible relatively high-redshift signal with $z > $ 3 if the drop is due to $\rm Ly\alpha$ absorption. So even if the burst brightness was close to the spectroscopic limit of BFOSC, we immediately acquired $3 \times 1800$ s exposure for spectroscopic observation about 0.3 day since the burst trigger.
Using $2\farcs$3 slit, the order-sorter filter 385LP and the grating of G4 with $1\times1$ binning, we luckily got spectral coverage of 3800 to 9000 {\AA} with 3 exposure. We processed the spectra using the standard {\em IRAF} data reduction and flux calibrated with standard star HD19445 \citep{Oke+1983} obtained by the same night with the same instrument setup. The reduced BFOSC spectrum is shown in the top panel of Fig.~\ref{fig:spec}. With the significant $\rm Ly\alpha $ drop, \ion{S}{2}, \ion{Si}{2}, the mixed \ion{C}{2} and the mixed \ion{C}{4} double lines, we measured the redshift of the burst as 4.61, which confirmed our previous hypothesis. The spectral identified redshift makes the most distant gamma-ray burst ever identified by BFOSC mounted on Xinglong 2.16 m.

About 0.6 day after the BAT trigger, we triggered ALFOSC mounted on the NOT for more detailed spectrum information. We used a $1\farcs$3 slit and a volume phase holographic grism with a fixed order-blocker filter called OG550, limiting the wavelength range from 5650 {\AA} to 10150 {\AA}. A $2\times2$ binning was chosen in order to improve the signal-to-noise ratio.
 
Using \emph{IRAF} standard spectrum processing tasks, the extracted spectrum is shown in the bottom panel of Fig.~\ref{fig:spec}. In the ALFOSC spectrum, we identify absorption lines from $\rm Ly \alpha$, \ion{N}{5}, \ion{S}{2}, \ion{Si}{2}, \ion{C}{2}, \ion{C}{2}$^*$ and \ion{C}{4}. The identified metal lines are marked in Fig.~\ref{fig:spec} and listed in Table~\ref{table:lines} with the measured equivalent widths (EW). We determine a redshift of $z = 4.615 \pm 0.001$, in full agreement with the BFOSC redshift.

\begin{figure}[htp]
\center
\includegraphics[width=0.45\textwidth]{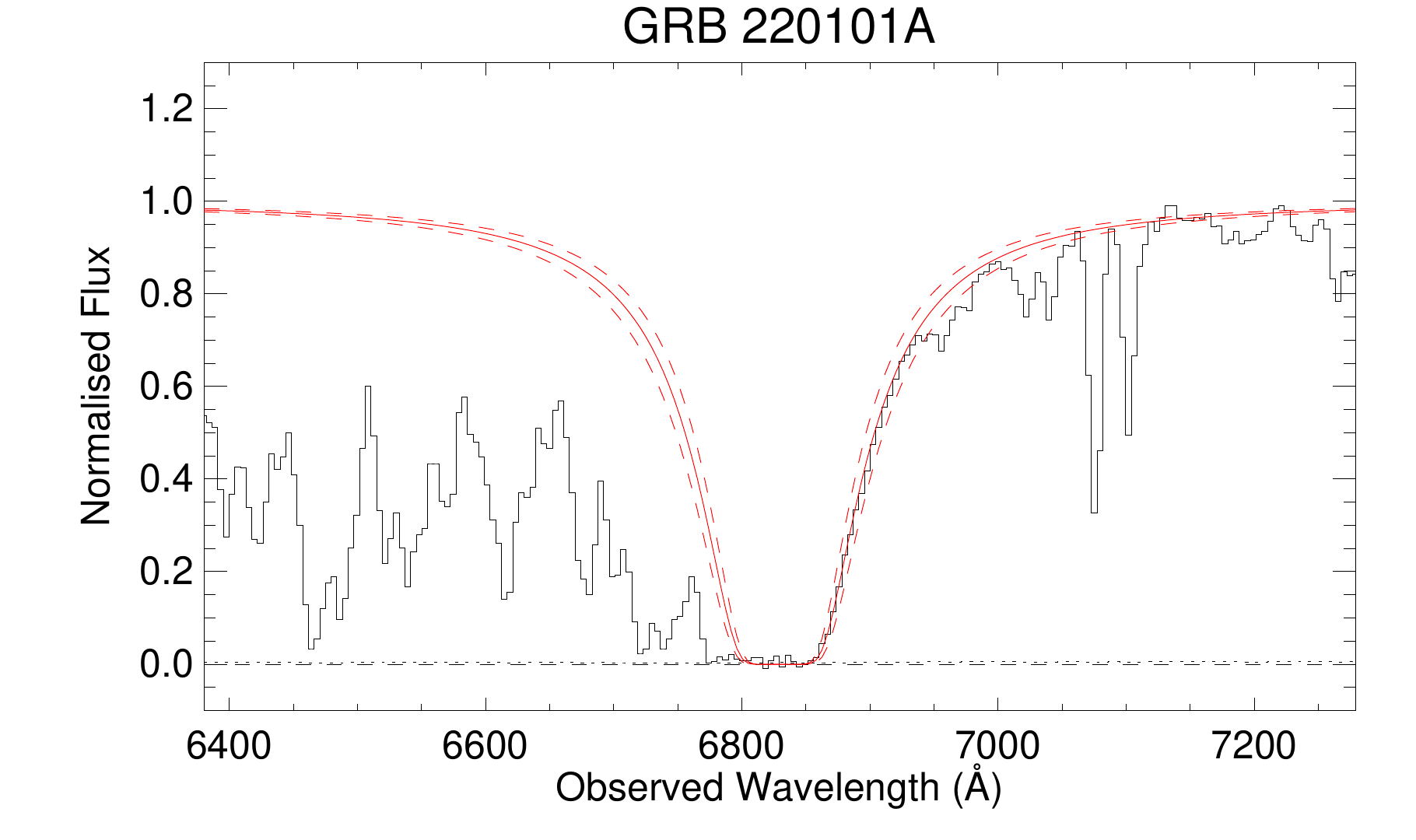}
\caption{The best fit of the $\rm Ly\alpha$ break. The hydrogen column density is $\rm log\ (N_{HI} / cm^{-2}) = 21.55 \pm 0.08$ at a redshift $z=4.615$.}
\label{fig:HI}
\end{figure}

\begin{figure}[htp]
\center
\includegraphics[width=0.45\textwidth]{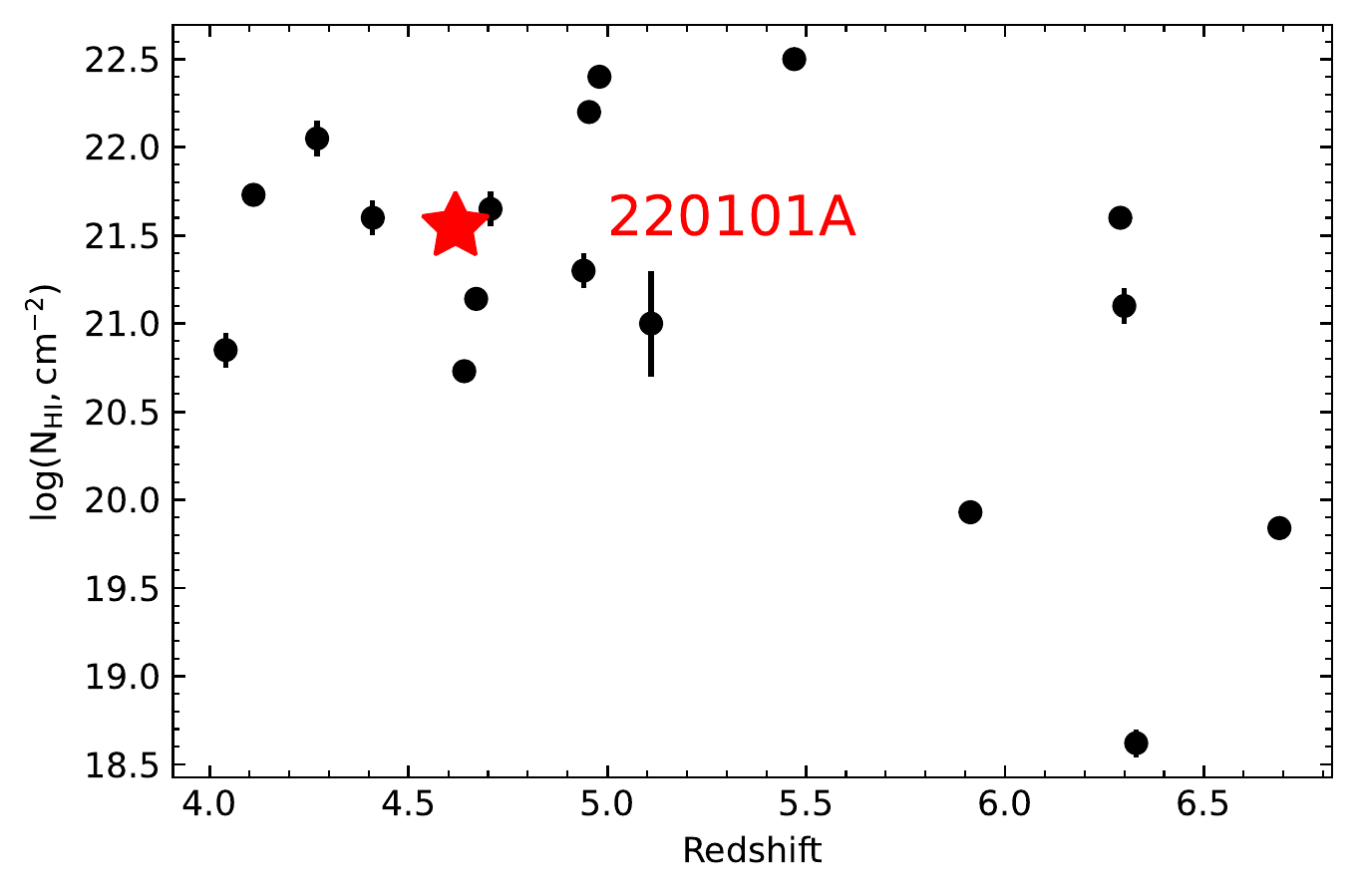}
\caption{The \ion{H}{1} column density of $z>4$ GRBs. The data points are collected from \cite{2013MNRAS.428.3590T,2014arXiv1405.7400C,2018ApJ...858...65L,2022arXiv221116524S}. }
\label{fig:HI_comp}
\end{figure}

Considering the higher S/N ratio of ALFOSC, the measured redshift $z=4.615$ is adopted as the redshift of the burst. We also fit the profile of $\rm Ly\alpha$, which is shown in Fig.~\ref{fig:HI} with the fit result of $\rm log\ (N_{HI} / cm^{-2}) = 21.55 \pm 0.08$. In Fig.~\ref{fig:HI_comp}, we compared the $\rm N_{HI}$ column density with other $z>4$ bursts. The hydrogen column density in the GRB host galaxy along the line of sight is similar to bursts at similar redshift.

\section{Multiband ANALYSIS AND EXTERNAL SHOCK MODELING} \label{sec:model}
\subsection{Mirror feature in prompt phase}

\begin{figure}[htp]
\center
\includegraphics[width=0.45\textwidth]{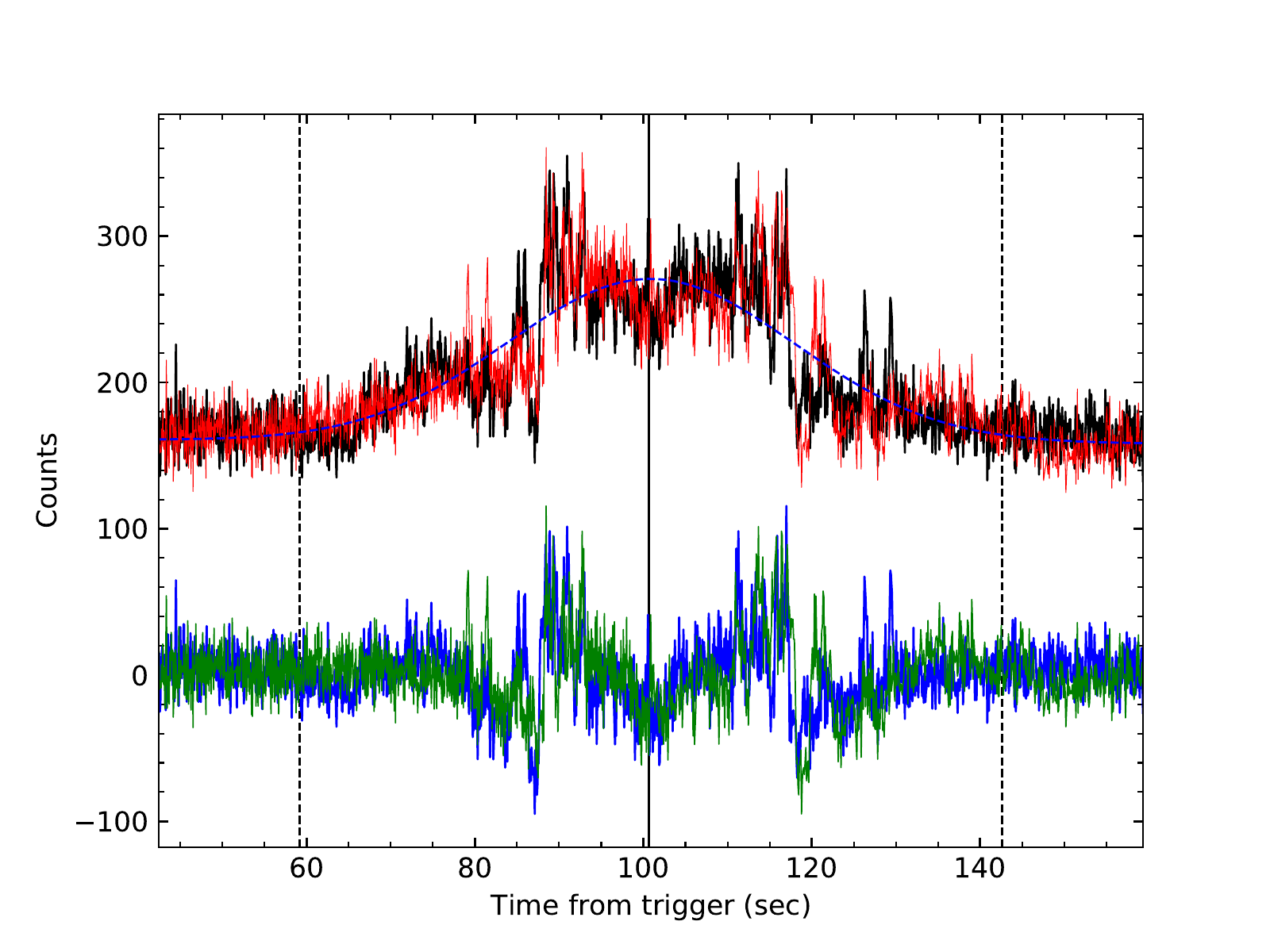}
\caption{Temporally symmetric model fit to GRB\,220101A light curve. Shown are the 64 ms counts data (black) obtained by NaI detectors n6 and n7 of Fermi/GBM, the fit to the monotonic components (blue dashed line), the time-reversed model (red), the residual (blue), the time-reversed residual (green), the duration window (vertical dashed lines), and the time of reflection (vertical solid line).
\label{fig:lcmirror}}
\end{figure}

\cite{2021ApJ...919...37H} studied a sample of BATSE GRBs and found that most GRB pulses can be characterized by a smooth single-peaked component coupled with a temporally symmetrical residual structure. 
Following their approach, we use the time-tagged event (TTE) data from Fermi Science Support Center's FTP site\footnote{\url{https://heasarc.gsfc.nasa.gov/FTP/fermi/data/gbm/bursts/}} and find that GRB\,220101A is a typical ``time symmetric" pulse. 
The result is shown in Fig.~\ref{fig:lcmirror}. The monotonic component is finally fitted by Gaussian model. The residual structure obtained by subtracting the monotonic component from data is highly symmetric. There are two main pulses, one is a typical pulse of fast rise and exponential decay (FRED) and the other is just its time symmetric form, with the symmetric time $t_{\rm 0;mirror}=100.7$ s, stretching parameter $s_{\rm mirror}=0.75$. The uncertainty of stretching parameter estimated by resampling the data shows $\sigma_{s, \rm mirror}=0.18$ \citep{2010arXiv1009.2755A}, satisfying the criterion $\sigma_{s,\rm mirror}<0.4$. Kinematic behaviors might explain this ``mirrored'' wavelike structures \citep{2018ApJ...863...77H,2019ApJ...883...70H}.

\subsection{Temporal analysis}

The obtained light curve is from $\sim$ 0.2 day to $\sim$ 30 day from optical to near-infrared, included \textit{g, r, i, z, J, H, K}. The single power-law (SPL) and broken power-law (BPL) are used to fit the decay index of \textit{r, i, z, J} and the X-ray band. The decay indexes of each band are listed in Table~\ref{tab:indices}. 
Note the earlier and later observations for \textit{r} band and \textit{i, J} band, respectively, 
we thus divide optics into three stages: 1) the shallow decay phase before $\sim 60$ ks, the decay index $\alpha_{O,1} \sim 0.7$, 2) the normal decay phase between $\sim 64$ ks and $\sim 1400$ ks, the decay index $\alpha_{O,2} \sim 1.3$, 3) the late decay after $\sim 1400$ ks and the decay index is about 3. However, we cannot determine the time of the second break and the decay index after the break for the very sparse points.  In our fit, the X-ray light curve can be divided into two stages, the early and late phase, with a break time at 65 ks. We also note that there is a significant flow drop at the last point of the X-ray, which may be a signal of accelerated decay at a later stage. This will be consistent with the three stages of optics. Our model fitting also supports the speculation above, which is shown in Fig.~\ref{fig:fit_model}.

\subsection{Afterglow SED analysis}

The analysis of Spectral Energy Distribution (SED) enables a deeper understanding of the afterglow evolution. In this study, we performed SED analysis on four epochs of XRT data (0.3-10 keV) obtained from  the online repository combined with optical data\footnote{\url{https://www.swift.ac.uk/xrt_spectra/}}, denoted from Epoch 1 to 4. The detail information of the epoch and optical data is listed in Table~\ref{tab:sed}. We utilized the \textit{Xspec} package (version 12.12) to fit the data of these epochs with single power-law model, accounting for the dust extinction of the host galaxy (Small Magellanic Cloud with $R_V=2.93$, \citealt{1992ApJ...395..130P}), photoelectric absorption of the host galaxy and the Milky Way, i.e. ``zdust*zphabs*phabs*powerlaw''. To enhance the signal-to-noise ratio, the X-ray data was rebinned using the ``grppha'' tool to ensure at least 20 photons per bin. The Galactic column density of Hydrogen ($\rm N_{H,Gal}=6.29\times 10^{20}$ cm$^{-2}$) was obtained from the Calculate Galactic NH tool\footnote{\url{https://www.swift.ac.uk/analysis/nhtot/}}.

\begin{figure}[thp]
\center
\includegraphics[width=0.45\textwidth]{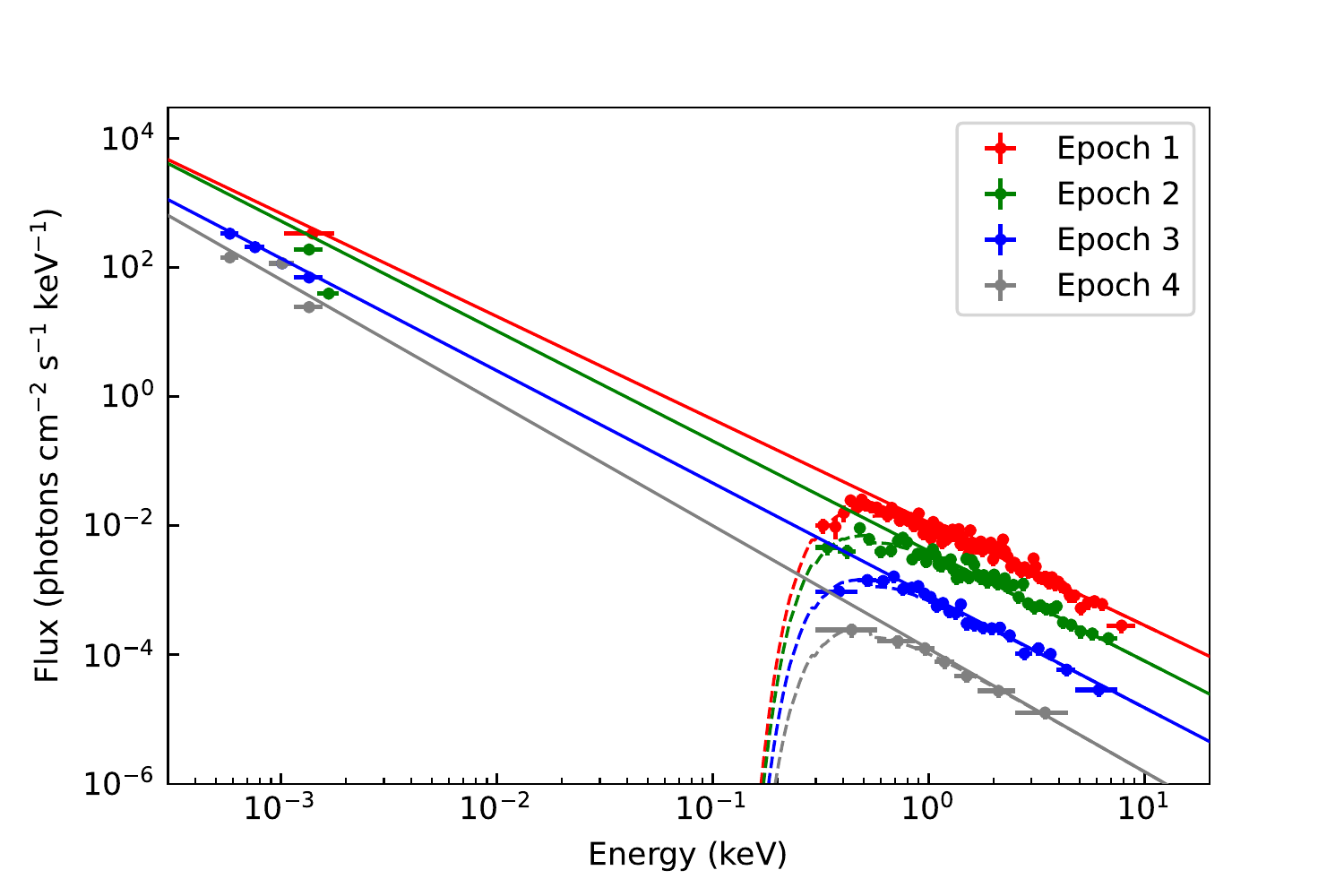}
\caption{The afterglow SED of GRB\,220101A at 30 ks (Epoch 1, red lines), 60 ks (Epoch 2, green lines), 150 ks (Epoch 3, blue lines) and 400 ks (Epoch 4, gray lines) from optical to X-ray. The optical multiband data are listed in Table~\ref{tab:sed}. The solid and dashed lines are the results of the single power-law and the model fit for each epoch.}
\label{fig:sed}
\end{figure}

Considering the abundant optical data available of Epoch 3, we performed SED fitting for this epoch first and obtained the host extinction $\rm E(B-V)=(1.69\pm 0.09)\times 10^{-2}$ and the host hydrogen column density $\rm N_{H,host}=(1.08\pm0.72)\times 10^{22}\ cm^{-2}$. We subsequently fixed the value above in the remaining fitting. The fit lines are shown in Fig.~\ref{fig:sed} and the photon indexes are also presented in Table~\ref{tab:sed}.

\subsection{External Shock Modeling}
In the standard external shock fireball model, the optical, near-infrared and X-ray emissions from afterglows can be understood with the synchrotron emission from the interaction between a relativistic jet and constant (ISM) or wind type external medium\citep{Rees1992,Meszaros1997,Sari1998,Zhang2018}. Electrons are believed to be accelerated at the shock front to a power-law distribution $N(\gamma_{\rm e}) \propto \gamma_{\rm e}^{-p}$. A fraction $\epsilon_{\rm e}$ of the shock energy is distributed into electrons, and a fraction $\epsilon_{\rm B}$ is in the magnetic field generated behind the shock. Accounting for the radiative cooling and the continuous injection of new accelerated electrons at the shock front, one expects a broken power-law energy spectrum of them, which leads to a multi-segment broken power-law radiation spectrum separated by three characteristic frequencies at any epoch: the synchrotron cooling frequency $\nu_{\rm c}$, the synchrotron frequency $\nu_{\rm m}$ defined by the minimum electron Lorentz factor, and the synchrotron self-absorption frequency $\nu_{\rm a}$ (below which the synchrotron photons are self-absorbed) \citep[for a review]{Gao+2013,Zhang2018}. Usually, the self-absorption frequency does not affect the X-ray and optical data at early epochs, and it mainly affects the low-frequency observations of afterglows. 

We assumed a constant external medium (ISM) and performed a multiband fit to GRB\,220101A afterglow data using the public python package \textit{afterglowpy}, which is an open-source numerical and analytic modelling tool to calculate the synchrotron light curve and spectrum from an external shock \citep{Ryan2020}. Such multiband modelling is helpful to constrain physical parameters associated with the afterglow. Markov chain Monte Carlo (MCMC) Ensemble sampler with python package \textit{emcee} \citep{2013PASP..125..306F} is adopted for multiband fitting to constraint the model parameters and associated errors.

Structured jet models (such as Top-Hat, Gaussian, Power-law, etc.) are involved in \textit{afterglowpy} to produce the light curves. In this paper, we took the Top-Hat jet type structure in the modelling to GRB\,220101A. Five free parameters are considered, i.e., the isotropic kinetic energy $E_{\rm K,iso}$, the half-width of the jet core $\theta_{\rm C}$, the number density of ISM medium $n_1$, the electron distribution power-law index $p$, the thermal energy fraction in electrons $\epsilon_{\rm e}$ and in magnetic field $\epsilon_{\rm B}$. The viewing angle $\theta_{\rm O}$ is fixed to 0 in the fit. 

\begin{figure*}[htp]
\center
\includegraphics[width=0.9\textwidth]{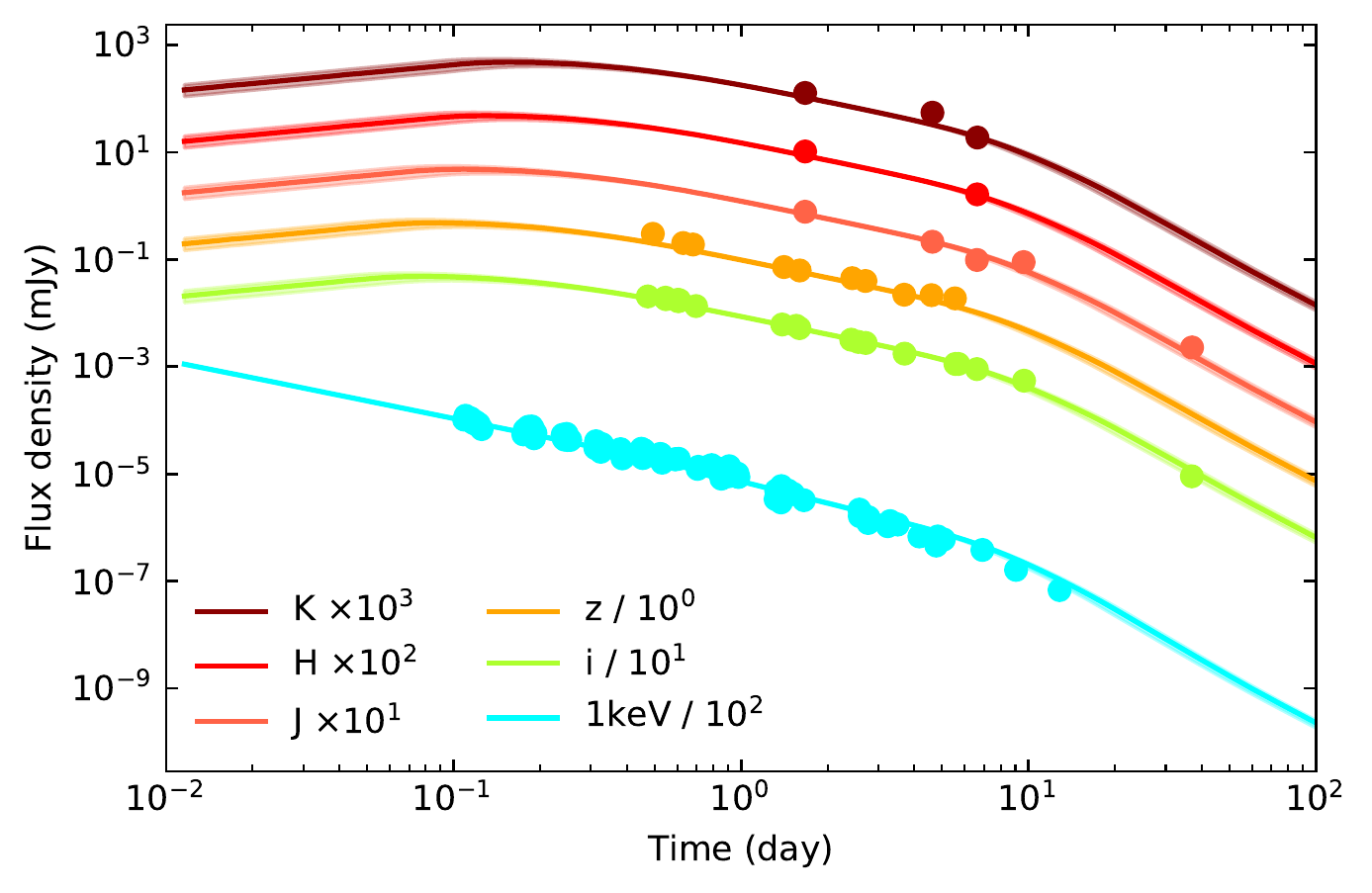}
\caption{Optical, near-infrared and X-ray data of GRB\,220101A along with the best fit afterglow modelling. The shaded region is the $2\sigma$ error region. X-ray flux density is converted from 0.3-10 keV to 1 kev followed \cite{2008ApJ...689.1161G}.}
\label{fig:fit_model}
\end{figure*}

\begin{figure*}[htp]
\center
\includegraphics[width=0.9\textwidth]{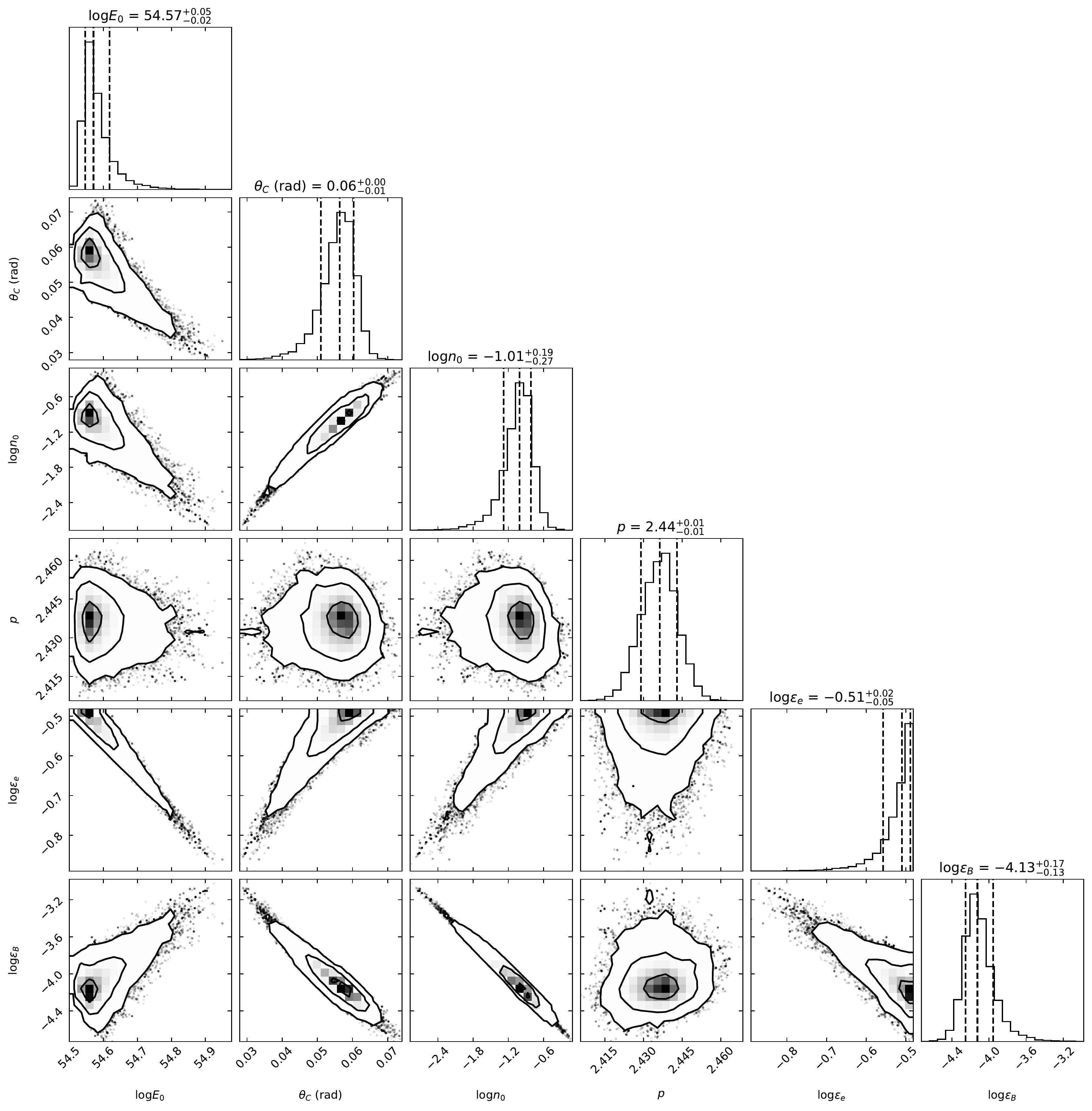}
\caption{Posterior distribution and parameter constraints obtained using multiband afterglow modelling of GRB\,220101A with \emph{afterglowpy}. The median values with the $1\sigma$ error regions are also shown in the one-dimensional probability distribution.}
\label{fig:model_corner}
\end{figure*}

We use Top-Hat jet model and performed a parameter search with 30 walkers over 15000 iterations, discarding the first 7500 as burn-in steps. The prior type and range for each model parameter are presented in Table~\ref{tab:fit}. In Fig.~\ref{fig:fit_model}, we have shown the optical and X-ray afterglow light curves of GRB\,220101A along with the best fit model. The contour plot of the model parameters is shown in Fig.~\ref{fig:model_corner}. The best fit of each parameter is given in Table~\ref{tab:fit} as: $E_{\rm K,iso}=3.52\times 10^{54}$ erg, $\theta_{\rm C}=3.43$ deg, $n_0 = 0.15 \ {\rm cm^{-3}}$, $p = 2.43$, $\epsilon_{\rm e}=0.326$ and $\epsilon_{\rm B}=5.86\times 10^{-5}$.

\section{Discussion} \label{sec:discussion}
As the \textit{r} and \textit{R} bands of burst are affected by the absorption of Lyman-$\alpha$ forest, both bands are excluded in our model fitting. According to the difference between the observed flux of \textit{R} band and the modeled \textit{R} band flux, we corrected the affected flux to the pseudo-real magnitude by adding $\Delta m = -1.1$ magnitude. In order to show the light curve of the whole period, we modified the light curve of the \textit{ white, r, J} bands to the \textit{R} band. The final whole period light curve compared with historic sample is shown in Fig.~\ref{fig:lc_comparision}. It is clear that the optical afterglow of GRB\,220101A is one of the most luminous ones ever observed, comparable to the high-\textit{z} burst GRB\,050904 \citep{Zou2006} and the Naked-eye burst GRB\,080319B \citep{2009ApJ...691..723B}, and exceeding even this well-know energetic event in some aspects.
The mirror feature in the prompt phase, indicting GRB\,220101A is a typical ``time symmetric'' pulse. The physics behind is still unknown.

After the deceleration time, the jet approaches the \citet{BM1976} self-similar evolution $\Gamma(t)\simeq (17E_{\rm K,iso}/1024\pi n_1 m_{\rm p}c^5t^3)^{1/8}$ and $R(t)\simeq (17E_{\rm K,iso} t/4\pi n_1 m_{\rm p}c)^{1/4}$ \citep{Lei2016}. Later, as the ejecta is decelerated to the post-jet-break phase at the time 
\begin{equation}
t_{\rm j}\simeq 0.6 {\rm day} \left(\frac{\theta_{\rm C}}{0.1 {\rm rad}}  \right)^{8/3}  \left(\frac{E_{\rm K,iso}}{10^{53} {\rm erg}} \right)^{1/3} n_1^{-1/3},
\end{equation}
when the $1/\Gamma$ cone becomes larger than $\theta_{\rm C}$.

As shown in Table \ref{tab:indices}, just after the shallow decay phase, a break appears in optical and X-ray at $\sim 6\times 10^5$ s. The change of temporal indices $\Delta \alpha \sim 0.7$, which is consistent with the prediction ($\Delta \alpha = 0.75$) from standard external shock model \citep{1999MNRAS.306L..39M,Gao+2013,Zhang2018}. Therefore, this break is likely the jet break. Using this jet break time, we can estimate the opening angle $\theta_{\rm C} \sim 3.8^\circ$ if we inset $E_{\rm K,iso}=3.52\times 10^{54}$ erg, $n_0=0.15\ \rm cm^{-3}$ and $t_{\rm j}\simeq 6\times 10^5/(1+z)$ s into the analytical expression Equation (1). As shown in Table \ref{tab:fit}, our numerical modeling gives the opening angle $\theta_{\rm C} \sim 3.43^\circ$, which is consistent with this analytical estimation.

From the observations, the isotropic $\gamma$-ray energy is $E_{\gamma, \rm iso} \simeq 3\times 10^{54}$ erg \citep{Mei2022}. From our modeling, we found the isotropic kinetic jet energy of $E_{\rm K,iso}=3.52\times 10^{54}$ erg. Therefore, the total jet energy is $E_{\rm total}=E_{\gamma, \rm iso} + E_{\rm K, iso} \simeq 6.52\times 10^{54}$ erg. The opening angle-corrected jet energy will be $E_{\rm j} \sim 6\times 10^{51}$ erg, which is well below the maximum rotational energy of  $3 \times 10^{52}$ erg \citep{2016PhR...621..127L} $-$ $7 \times 10^{52}$ erg \citep{2009A&A...502..605H} for a standard neutron star with mass $M\sim1.4\,M_\sun$. Therefore, our data do not require a black hole as the central engine of this GRB.

\section{Summary} \label{sec:summary}
We present our optical and near-infrared observations of the relatively high-\textit{z} `New Year's Burst' GRB\,220101A. With the optical spectrum obtained by ALFOSC and BFOSC, we measure the redshift by the significant metal lines. Combine with our multiband data and X-ray light curve obtained by XRT, we perform multiband fit with the Python package \emph{afterglowpy}. Our conclusions are summarized as follows:

1. The redshift of the burst is $z=4.615\pm0.001$. The \ion{H}{1} column density in the GRB host galaxy along the line of sight is $\rm log\ (N_{HI} / cm^{-2}) = 21.55 \pm 0.08$, consistent with nearby GRB host galaxies.

2. A mirror feature is found in the prompt phase. The physical origin is unclear.

3. Comparison of a large sample of GRB optical afterglow light curves shifted in time and flux to a common redshift of $z = 1$, it is clear that GRB\,220101A is one of the most luminous GRBs ever observed.

4. The multiband afterglow data (optical, near-infrared and X-ray) can be interpreted with the standard external shock model. From the observations and modeling, we found that the total jet energy is $E_{\rm total}=E_{\gamma, \rm iso} + E_{\rm K, iso} \simeq 6.52\times 10^{53}$ erg.

5. The breaks at a few $\times 10^5$ s in both X-ray and optical bands are roughly consistent with the jet break, revealing an opening angle $\sim 3.43^\circ$. The opening angle-corrected jet energy will be $E_{\rm j} \sim 6\times 10^{51}$ erg.

6. We find that the fit parameters are consistent with the typical afterglow parameters of other well studied GRBs.

\section*{acknowledgments}

We acknowledge the support of the staff of the Xinglong 2.16m telescope, NOT, NEXT, CAHA 2.2m, Tautenburg, TNG and GTC.
The data presented here were obtained in part with ALFOSC, which is provided by the Instituto de Astrofisica de Andalucia (IAA) under a joint agreement with the University of Copenhagen and NOT. 
This research has made use of the Spanish Virtual Observatory (http://svo.cab.inta-csic.es) supported by the MINECO/FEDER through grant AyA2017-84089.7. 
This work was also partially supported by the Open Project Program of the Key Laboratory of Optical Astronomy, National Astronomical Observatories, Chinese Academy of Sciences. 
Based on observations made with the Gran Telescopio Canarias (GTC), installed at the Spanish Observatorio del Roque de los Muchachos of the Instituto de Astrofísica de Canarias, on the island of La Palma, under program ID GTCMULTIPLE2H-21B.
This research is based on observations made with the NASA/ESA Hubble Space Telescope obtained from the Space Telescope Science Institute, which is operated by the Association of Universities for Research in Astronomy, Inc., under NASA contract NAS 5–26555. These observations are associated with program 16838.
Based on observations made with the NASA/ESA Hubble Space Telescope, and obtained from the Hubble Legacy Archive, which is a collaboration between the Space Telescope Science Institute (STScI/NASA), the Space Telescope European Coordinating Facility (ST-ECF/ESAC/ESA) and the Canadian Astronomy Data Centre (CADC/NRC/CSA).
This work is supported by the National Key R\&D Program of China (Nos. 2020YFC2201400), the National Natural Science Foundation of China under grants U2038107,and U1931203. 
W.H.Lei. acknowledges support by the science research grants from the China Manned Space Project with NO.CMS-CSST-2021-B11. J.P.U.F. acknowledgs support from the Carlsberg foundation. 
Data resources are supported by China National Astronomical Data Center (NADC) and Chinese Virtual Observatory (China-VO). The Cosmic Dawn Center (DAWN) is funded by the Danish National Research Foundation under grant No.140. 
This work is supported by Astronomical Big Data Joint Research Center, co-founded by National Astronomical Observatories, Chinese Academy of Sciences and Alibaba Cloud. We acknowledge the use of public data from the {\em Swift} data archive.

\software{
Afterglowpy \citep{Ryan2020},
Astrometry.net \citep{Lang+2010},
Astropy \citep{2022ApJ...935..167A},
EMCEE \citep{2013PASP..125..306F},
IRAF \citep{Tody+1986},
Matplotlib \citep{Hunter:2007},
Numpy \citep{harris2020array},
Photutils \citep{2021zndo...5525286B},
PSFEx \citep{2011ASPC..442..435B},
Python \citep{10.5555/1593511},
Scipy \citep{2020SciPy-NMeth},
Source Extractor \citep{BertinArnouts+1996}
}

\bibliography{ref0}{}
\bibliographystyle{aasjournal}

\begin{center}
\begin{longtable}{c c c c c}
\caption{The photometric results of our observations combined with collected GCN results. $\Delta T$ is the exposure medium time after the BAT trigger. Magnitude in AB system is \emph{not} corrected for Galactic Extinction, which is E(B-V) = 0.05 \citep{2011ApJ...737..103S}. References: (1) this work, (2) \citealt{2022GCN.31357....1P}, (3) \citealt{2022GCN.31425....1P}} \label{tab:phot} \\
\hline \multicolumn{1}{c}{\textbf{$\Delta T$(day)}} & \multicolumn{1}{c}{\textbf{Telescope/Instrument}} & \multicolumn{1}{c}{\textbf{Filter}} & \multicolumn{1}{c}{\textbf{Mag(AB)}} & \multicolumn{1}{c}{\textbf{Ref.}} \\ \hline 
\endfirsthead

\multicolumn{5}{c}%
{{\bfseries \tablename\ \thetable{} -- continued from previous page}} \\
\hline \multicolumn{1}{c}{\textbf{$\Delta T$(day)}} & \multicolumn{1}{c}{\textbf{Telescope/Instrument}} & \multicolumn{1}{c}{\textbf{Filter}} & \multicolumn{1}{c}{\textbf{Mag(AB)}} & \multicolumn{1}{c}{\textbf{Ref.}} \\ \hline 
\endhead

\hline \multicolumn{5}{r}{{Continued on next page}} \\ 
\endfoot

\hline \hline
\endlastfoot
0.219 & Xinglong 2.16m/BFOSC & $I$ & 17.76 $\pm$ 0.01 & (1) \\
0.223 & Xinglong 2.16m/BFOSC & $R$ & 18.86 $\pm$ 0.02 & (1) \\
0.227 & Xinglong 2.16m/BFOSC & $V$ & 19.88 $\pm$ 0.06 & (1) \\
0.231 & Xinglong 2.16m/BFOSC & $B$ & $>$21.9   & (1) \\
2.203 & Xinglong 2.16m/BFOSC & $R$ & 21.26 $\pm$ 0.08 & (1) \\
2.227 & Xinglong 2.16m/BFOSC & $I$ & 19.95 $\pm$ 0.03 & (1) \\
2.255 & Xinglong 2.16m/BFOSC & $V$ & $>$22.0  & (1) \\
4.219 & Xinglong 2.16m/BFOSC & $I$ & 20.80 $\pm$ 0.04 & (1) \\
4.255 & Xinglong 2.16m/BFOSC & $R$ & 22.28 $\pm$ 0.13 & (1) \\
7.221 & Xinglong 2.16m/BFOSC & $I$ & 20.94 $\pm$ 0.06 & (1) \\
0.327 & NEXT & $r$ & 19.30 $\pm$ 0.06 & (1) \\
0.436 & NEXT & $g$ & $>$21.6  & (1) \\
0.454 & NEXT & $r$ & 19.52 $\pm$ 0.06 & (1) \\
0.474 & NEXT & $i$ & 18.22 $\pm$ 0.03 & (1) \\
0.493 & NEXT & $z$ & 17.77 $\pm$ 0.08 & (1) \\
1.390 & NEXT & $i$ & 19.53 $\pm$ 0.08 & (1) \\
1.410 & NEXT & $z$ & 19.32 $\pm$ 0.17 & (1) \\
1.443 & NEXT & $r$ & 21.12 $\pm$ 0.17 & (1) \\
2.415 & NEXT & $i$ & 20.24 $\pm$ 0.16 & (1) \\
2.396 & NEXT & $r$ & 21.40 $\pm$ 0.25 & (1) \\
2.415 & NEXT & $i$ & 20.24 $\pm$ 0.16 & (1) \\
2.435 & NEXT & $z$ & 19.84 $\pm$ 0.22 & (1) \\
0.544 & CAHA 2.2m/CAFOS & $i$ & 18.30 $\pm$ 0.03 & (1) \\
0.546 & CAHA 2.2m/CAFOS & $i$ & 18.28 $\pm$ 0.03 & (1) \\
0.548 & CAHA 2.2m/CAFOS & $i$ & 18.35 $\pm$ 0.03 & (1) \\
0.549 & CAHA 2.2m/CAFOS & $r$ & 19.77 $\pm$ 0.06 & (1) \\
0.551 & CAHA 2.2m/CAFOS & $r$ & 19.80 $\pm$ 0.06 & (1) \\
0.552 & CAHA 2.2m/CAFOS& $r$ & 19.71 $\pm$ 0.05 & (1) \\
0.556 & CAHA 2.2m/CAFOS& $g$ & 22.19 $\pm$ 0.13 & (1) \\
0.592 & CAHA 2.2m/CAFOS& $r$ & 19.84 $\pm$ 0.06 & (1) \\
0.596 & CAHA 2.2m/CAFOS& $r$ & 19.79 $\pm$ 0.04 & (1) \\
0.598 & CAHA 2.2m/CAFOS& $r$ & 19.85 $\pm$ 0.04 & (1) \\
0.600 & CAHA 2.2m/CAFOS& $r$ & 19.83 $\pm$ 0.04 & (1) \\
0.601 & CAHA 2.2m/CAFOS& $i$ & 18.38 $\pm$ 0.02 & (1) \\
0.604 & CAHA 2.2m/CAFOS& $i$ & 18.44 $\pm$ 0.02 & (1) \\
0.605 & CAHA 2.2m/CAFOS& $i$ & 18.40 $\pm$ 0.02 & (1) \\
0.607 & CAHA 2.2m/CAFOS& $i$ & 18.43 $\pm$ 0.03 & (1) \\
0.609 & CAHA 2.2m/CAFOS& $i$ & 18.39 $\pm$ 0.03 & (1) \\
0.611 & CAHA 2.2m/CAFOS& $i$ & 18.42 $\pm$ 0.03 & (1) \\
1.545 & CAHA 2.2m/CAFOS& $r$ & 21.02 $\pm$ 0.05 & (1) \\
1.556 & CAHA 2.2m/CAFOS& $i$ & 19.59 $\pm$ 0.03 & (1) \\
2.556 & CAHA 2.2m/CAFOS& $i$ & 20.34 $\pm$ 0.04 & (1) \\
6.597 & CAHA 2.2m/CAFOS& $r$ & $>$22.7   & (1) \\
6.611 & CAHA 2.2m/CAFOS& $i$ & 21.61 $\pm$ 0.17 & (1) \\
0.658 & NOT/ALFOSC & $r$ & 19.87 $\pm$ 0.01 & (1) \\
5.687 & NOT/ALFOSC & $r$ & 22.74 $\pm$ 0.13 & (1) \\
5.699 & NOT/ALFOSC & $i$ & 21.37 $\pm$ 0.09 & (1) \\
9.644 & NOT/ALFOSC & $i$ & 22.15 $\pm$ 0.09 & (1) \\
0.710 & Tautenburg 1.34m/TAUKAM & $r$ & 19.86 $\pm$ 0.13 & (1) \\ 
1.666 & Tautenburg 1.34m/TAUKAM& $g$ & $>$20.96 & (1) \\ 
5.545 & Tautenburg 1.34m/TAUKAM & $z$ & 20.78 $\pm$ 0.16 & (1) \\ 
5.565 & Tautenburg 1.34m/TAUKAM & $i$ & 21.37 $\pm$ 0.10 & (1) \\ 
5.599 & Tautenburg 1.34m/TAUKAM & $r$ & 22.81 $\pm$ 0.24 & (1) \\ 
5.634 & Tautenburg 1.34m/TAUKAM & $g$ & $>$23.8 & (1) \\ 
1.67 & TNG/NICS & $J$ & 19.21 $\pm$ 0.07 & (1) \\ 
1.67 & TNG/NICS & $H$ & 18.89 $\pm$ 0.06 & (1) \\
1.67 & TNG/NICS & $Ks$ & 18.65 $\pm$ 0.07 & (1) \\
4.63 & TNG/NICS & $J$ & 20.61 $\pm$ 0.10 & (1) \\
4.63 & TNG/NICS & $Ks$ & 19.57 $\pm$ 0.10 & (1) \\
9.61 & TNG/NICS & $J$ & 21.56 $\pm$ 0.25 & (1) \\
6.604 & GTC/EMIR & $Y$ & 22.40 $\pm$ 0.16 & (1) \\
6.614 & GTC/EMIR & $J$ & 21.46 $\pm$ 0.14 & (1) \\ 
6.623 & GTC/EMIR & $H$ & 20.89 $\pm$ 0.12 & (1) \\
6.637 & GTC/EMIR & $Ks$ & 20.73 $\pm$ 0.08 & (1) \\
36.97 & HST/WFC3 & F775W & 26.61 $\pm$ 0.08 & (1)\\
37.04 & HST/WFC3 & F125W & 25.54 $\pm$ 0.05 & (1)\\
0.624 & LT/IO:O & $g$ & 21.97 $\pm$ 0.1 & (2) \\
0.625 & LT/IO:O & $R$ & 19.84 $\pm$ 0.03 & (2) \\
0.627 & LT/IO:O & $I$ & 18.48 $\pm$ 0.03 & (2) \\
0.629 & LT/IO:O & $z$ & 18.20 $\pm$ 0.03 & (2) \\
0.675 & LT/IO:O & $g$ & 22.21 $\pm$ 0.13 & (2) \\
0.677 & LT/IO:O & $r$ & 19.92 $\pm$ 0.03 & (2) \\
0.678 & LT/IO:O & $I$ & 18.58 $\pm$ 0.03 & (2) \\
0.680 & LT/IO:O & $z$ & 18.27 $\pm$ 0.03 & (2) \\
1.592 & LT/IO:O & $r$ & 21.17 $\pm$ 0.06 & (3) \\
1.596 & LT/IO:O & $i$ & 19.70 $\pm$ 0.04 & (3) \\
1.599 & LT/IO:O & $z$ & 19.48 $\pm$ 0.05 & (3) \\
2.707 & LT/IO:O & $z$ & 19.97 $\pm$ 0.12 & (3) \\
2.712 & LT/IO:O & $i$ & 20.39 $\pm$ 0.08 & (3) \\
2.715 & LT/IO:O & $r$ & 21.58 $\pm$ 0.17 & (3) \\
3.693 & LT/IO:O & $z$ & 20.61 $\pm$ 0.12 & (3) \\
3.698 & LT/IO:O & $r$ & 22.31 $\pm$ 0.17 & (3) \\
3.703 & LT/IO:O & $i$ & 20.89 $\pm$ 0.09 & (3) \\
4.596 & LT/IO:O & $z$ & 20.63 $\pm$ 0.11 & (3) \\
4.601 & LT/IO:O & $i$ & 21.25 $\pm$ 0.11 & (3) \\
4.605 & LT/IO:O & $r$ & 23.13 $\pm$ 0.34 & (3) \\
6.702 & LT/IO:O & $i$ & 21.34 $\pm$ 0.19 & (3) \\
\end{longtable}
\end{center}

\begin{table}[htbp]
	\begin{center}{
			\caption{List of features of spectra and their equivalent widths.}
			\label{table:lines}
			\begin{tabular}{cccc} \hline\hline
				$\rm \lambda_{obs}$(\AA) & Feature(\AA) & $z$ & $\rm EW_{obs}$(\AA) \\
				\hline
				\hline
				\multicolumn{4}{c}{ALFOSC}\\
				\hline
6952.5 & NV$\lambda\lambda$1238.2 & 4.615 & 0.64$\pm$0.26  \\
7019.76 & SII$\lambda\lambda$1250.0 & 4.616 & 0.91$\pm$0.48  \\
7037.41 & SII$\lambda\lambda$1253.2 & 4.616 & 0.91$\pm$0.48  \\
7072.43 & SiII$\lambda\lambda$1259.8 & 4.614 & 4.87$\pm$1.32  \\
7098.67 & SiII$\lambda\lambda$1264.2 & 4.615 & 3.68$\pm$1.46  \\
7315.86 & OI/SiII$\lambda\lambda$1302.7 & 4.616 & 9.01$\pm$1.84  \\
7492.53 & CII/CII*$\lambda\lambda$1334.5 & 4.614 & 8.31$\pm$1.58  \\
7821.88 & SiII$\lambda\lambda$1393.2 & 4.614 & 3.16$\pm$1.28  \\
7872.56 & SiII$\lambda\lambda$1402.2 & 4.614 & 2.55$\pm$1.28  \\
8567.87 & SiII$\lambda\lambda$1526.1 & 4.614 & 6.83$\pm$1.53  \\
8605.41 & SiII*$\lambda\lambda$1532.9 & 4.614 & 3.3$\pm$1.4  \\
8694.91 & CIV/CIV$\lambda\lambda$1548.9 & 4.614 & 10.02$\pm$1.76  \\
				\hline
			\end{tabular}
		}
	\end{center}
\end{table}

\begin{table}[htbp]
\begin{center}
\caption{List of optical, near-infrared and X-ray light curve decay indices.}
\label{tab:indices}
\begin{tabular}{ccccccc}
\hline\hline
            Band & Model & $\alpha$ & $\alpha_1$  &$\alpha_2$ & $t_b$ (ks) & $\chi^2$/dof \\  
    \hline
        \multirow{2}{*}{\emph{r}}   & SPL & $1.17 \pm 0.04$ & $\cdot\cdot\cdot$ & $\cdot\cdot\cdot$ & $\cdot\cdot\cdot$  &  79.3/21\\
           & BPL & $\cdot\cdot\cdot$ & $0.70 \pm 0.10$ & $1.36 \pm 0.08$ & $63.8\pm9.7$ & 20.5/20 \\
        \hline
        \multirow{2}{*}{\emph{i}}   & SPL & $1.21 \pm 0.02$ & $\cdot\cdot\cdot$ & $\cdot\cdot\cdot$ & $\cdot\cdot\cdot$ & 39.9/21\\
           & BPL & $\cdot\cdot\cdot$ & $0.63 \pm 0.21$ & $1.27 \pm 0.03$ & $55.5\pm7.1$ &  16.0/20\\
        \hline
        \multirow{2}{*}{\emph{z}}   & SPL & $1.19 \pm 0.04$ & $\cdot\cdot\cdot$ & $\cdot\cdot\cdot$ & $\cdot\cdot\cdot$ & 8.5/6 \\
           & BPL & $\cdot\cdot\cdot$ & $1.36 \pm 0.15$ & $0.97 \pm 0.06$ & $114.8\pm40.8$ & 1.9/5 \\
        \hline
        \multirow{2}{*}{\emph{J}}   & SPL & $1.84 \pm 0.11$ & $\cdot\cdot\cdot$ & $\cdot\cdot\cdot$ & $\cdot\cdot\cdot$ & 46.6/3\\
           & BPL & $\cdot\cdot\cdot$ & $1.27 \pm 0.01$ & $3.52 \pm 0.06$   & $1393.6\pm16.6$ & 0.1/3\\
        \hline
        \multirow{2}{*}{X-ray}    & SPL & $1.37 \pm 0.02$ & $\cdot\cdot\cdot$ & $\cdot\cdot\cdot$ & $\cdot\cdot\cdot$ & 644.2/276\\
             & BPL & $\cdot\cdot\cdot$ & $1.01 \pm 0.04$ & $1.75 \pm 0.05$   & $64.9\pm6.5$   & 390.3/275\\
    \hline
\end{tabular}
\end{center}
\end{table}

\begin{table*}[htbp]
\begin{center}
\caption{SED data at four different epochs with the best fit indexes. 
The photometric data listed here are corrected for Galactic extinction.
}
\label{tab:sed}
\begin{tabular}{ccccc}
\hline\hline
                 Time & Epoch 1 (20 ks) & Epoch 2 (54 ks)  & Epoch 3 (140 ks)  & Epoch 4 (400 ks)  \\  
    \hline
       
               \emph{B} & $>$ 21.7  & $\cdot\cdot\cdot$ & $\cdot\cdot\cdot$ & $\cdot\cdot\cdot$ \\ 
               \emph{V} & $19.72\pm0.06$  & $\cdot\cdot\cdot$ & $\cdot\cdot\cdot$ & $\cdot\cdot\cdot$ \\ 
               \emph{R} & $18.74\pm0.02$  & $\cdot\cdot\cdot$ & $\cdot\cdot\cdot$ & $\cdot\cdot\cdot$ \\ 
               \emph{I} & $17.67\pm0.01$  & $\cdot\cdot\cdot$ & $\cdot\cdot\cdot$ & $\cdot\cdot\cdot$ \\ 
               \emph{g} & $\cdot\cdot\cdot$  & $21.78\pm0.10$ & $\cdot\cdot\cdot$ & $\cdot\cdot\cdot$ \\ 
               \emph{r} & $\cdot\cdot\cdot$  & $19.74\pm0.01$ & $21.04\pm0.06$ & $\cdot\cdot\cdot$ \\ 
               \emph{i} & $\cdot\cdot\cdot$  & $18.32\pm0.03$ & $19.60\pm0.04$ & $21.15\pm	0.11$ \\ 
               \emph{z} & $\cdot\cdot\cdot$  & $18.13\pm0.03$ & $19.41\pm0.05$ & $20.56\pm	0.11$ \\ 
               \emph{J} & $\cdot\cdot\cdot$  & $\cdot\cdot\cdot$ & $19.17\pm0.07$ & $20.57\pm0.10$ \\ 
               \emph{H} & $\cdot\cdot\cdot$  & $\cdot\cdot\cdot$ & $18.86\pm0.06$ & $\cdot\cdot\cdot$ \\ 
               \emph{Ks} & $\cdot\cdot\cdot$  & $\cdot\cdot\cdot$ & $18.63	\pm 0.07$ & $19.55\pm0.10$ \\ 
               \hline
               Photon Index & $-1.60\pm0.01$ & $-1.70\pm0.01$ & $-1.74\pm0.01$ & $-1.91\pm0.01$ \\
    \hline
\end{tabular}
\end{center}
\end{table*}

\begin{table}[htbp]
\begin{center}
\caption{The input parameters, prior type, prior range, best-fit value of multiband modelling of GRB\,220101A performed with \emph{afterglowpy}.}
\label{tab:fit}
\begin{tabular}{lrrr}
\hline\hline
     Parameters &  Prior Type & Prior Range  & Best fit  \\  
    \hline
    $E_{\rm K, iso}$ (erg) & log flat & $10^{52}-6\times 10^{56}$ & $3.52\times10^{54}$  \\ 
    $\theta_{\rm C}$ (rad) & flat & $0-0.2$ & 0.06  \\ 
    $n_{0}\rm\ (cm^{-3})$   & log flat & $10^{-6}-10^5$ & 0.15\\ 
    $p$   & flat & $2.3-2.7$ & 2.43 \\ 
    $\epsilon_{\rm e}$   & log flat & $10^{-6}-0.33$ &  $3.26\times10^{-1}$  \\ 
    $\epsilon_{\rm B}$   & log flat & $10^{-6}-0.33$ &  $5.86\times10^{-5}$  \\ 
    \hline
\end{tabular}
\end{center}
\end{table}

\end{document}